\documentclass[fleqn,usenatbib]{mnras}

\usepackage{newtxtext,newtxmath}

\usepackage[T1]{fontenc}
\usepackage{xcolor}
\usepackage{colortbl}
\usepackage{ulem}

\newcommand{\JLM}[1]{#1}
\newcommand{\JL}[1]{#1}

\DeclareRobustCommand{\VAN}[3]{#2}
\let\VANthebibliography\thebibliography
\def\thebibliography{\DeclareRobustCommand{\VAN}[3]{##3}\VANthebibliography}

\usepackage{graphicx}	
\usepackage{comment}

\usepackage{tabulary}
\usepackage{booktabs}

\newcommand{\lsi}{LS~I~+61\textdegree 303}

\title[Rapid variability in \lsi{}]{Rapid X-ray variability of the gamma-ray binary \lsi{}}
\author[J. López-Miralles et al.]{
J. López-Miralles,$^{1,4}$\thanks{E-mail: jose.lopez-miralles@uv.es}
Sara E. Motta,$^{2,3}$
S. Migliari$^{4}$ and
F. Jaron$^{5
}$
\\
$^{1}$Departament d’Astronomía i Astrofísica, Universitat de València, Dr. Moliner 50, 46100, Burjassot (València), Spain\\
$^{2}$INAF–Osservatorio Astronomico di Brera, via E. Bianchi 46, 23807 Merate (LC), Italy\\
$^{3}$Department of Physics, Astrophysics, University of Oxford, Denys Wilkinson Building, Keble Road, OX1 3RH Oxford, UK\\
$^{4}$Aurora Technology for the European Space Agency, ESAC/ESA, Camino Bajo del Castillo s/n, Urb. Villafranca del Castillo,\\ 
28691 Villanueva de la Cañada, Madrid, Spain\\
$^{5}$ Technische Universit\"at Wien, Wiedner Hauptstra\ss{}e 8-10, A-1040 Wien, Austria\\
}

\date{Accepted XXX. Received YYY; in original form ZZZ}

\pubyear{2023}

\begin{document}
\label{firstpage}
\pagerange{\pageref{firstpage}--\pageref{lastpage}}

\maketitle

\begin{abstract}
The gamma-ray binary \lsi{} has been widely monitored at different wavelengths since its discovery more than sixty years ago. However, the nature of the compact object and the peculiar behavior of the system are still largely debated. Aimed at investigating the rapid X-ray variability of \lsi{}, we have analysed all the archival RXTE/PCA data of the source, taken between 1996 and 2011. The timing analysis yields a periodicity of $P\sim 26.6\pm 0.3$~days, which is statistically compatible with several periodicities reported in the literature for \lsi{}. Using this period, we performed a data phase-resolved analysis to produce a set of phase-bin-averaged energy spectra and power density spectra. These power density spectra are dominated by weak red noise below $0.1$~Hz, and show no signal above this frequency. The amplitude of the red noise varies mildly with the phase, and shows a maximum that coincides with a dip of the X-ray flux and a softer photon index. Aside from low-frequency noise, this analysis does not provide any statistically significant periodic or quasi-periodic timing feature in the RXTE/PCA data of \lsi{}.

\end{abstract}




\section{Introduction}
\label{sec:introduction}


The high-mass X-ray binary (HMXB) \lsi{} consists of a compact object with an eccentric orbit ($e\approx 0.7$) around a rapidly rotating B0 Ve type star \citep{casares05}, whose orbital period is \JL{$P_{1} = 26.496 \pm 0.0028$~days} \citep{Gregory2002}. The nature of the compact object (CO) in \lsi{} is still unclear. Since optical observations are not accurate enough to place strong constraints on the mass function of the object, which is in this case affected by large statistical uncertainties \citep{casares05}, some authors argue that the central engine of the binary is a stellar-mass black hole \citep[BH; see e.g.,][]{punsly99,massi17}, while others suggest that the system contains a neutron star \citep[NS; see e.g.,][]{maraschi81,torres12}.

\lsi{} has been detected in a wide range of wavelengths, from radio \citep[i.e., non-thermal synchroton, see e.g.,][]{Gregory1978} to X-rays \citep[see e.g.,][]{bignami81}, as well as at high-energies, at GeV \citep{Abdo2009}, and very high-energies, at TeV \citep{Albert2006}. It is therefore  one of the very few $\gamma$-ray emitting binaries \citep[see, e.g.,][and references therein]{mirabel07,chernyakova20}, together with other canonical systems as LS 5039, HESS J0632+057 or 1FGL J1018.6-5856. 

In the X-ray energy range, \lsi{} is a weak source, with an average luminosity of $L_x\sim 10^{33}$~erg/s \citep{bignami81}. The Rossi X-ray Timing Explorer (RXTE) satellite \citep{harrison00,greiner01,leahy01,smith09,paredes97,Torres10,Li11}, as well as instruments on other missions, such as Swift-XRT \citep{esposito07} or INTEGRAL-IBIS/ISGRI \citep{ubertini09,zhang10}, allowed to perform long-term X-ray monitoring of \lsi{}, while soft X-ray pointed observations performed by XMM-Newton \citep{sidoli06}, Chandra \citep{paredes07,karg14}, ASCA \citep{leahy97}, ROSAT \citep{taylor96} and Einstein \citep{bignami81} were in general too short to cover a full single orbit. In 1996, the source was intensively observed by RXTE for one entire orbital cycle \citep{harrison00}. Other authors \citep{greiner01,neronov07} analyzed the same data set and found that the energy spectrum could be fitted properly by a simple, relatively hard, absorbed power law, suggesting an underlying non-thermal X-ray emission mechanism. \cite{smith09} analyzed five months of RXTE/PCA observations and found that the light-curve showed a two-peak profile in the 2-10 keV band, where the flux and the photon index were anti-correlated. Similar results were previously reported by \cite{paredes97}, but using RXTE/ASM data. Later on, \cite{Torres10} and \cite{Li11} considered respectively 35 and 42 full cycles of the source orbital motion. They showed that the orbital profile was not stable (meaning that the phase of the light-curve peak varied over time) and reported a strong anti-correlation between the X-ray flux and the source photon index. \JL{Together with the X-ray flux modulation, at radio wavelenghts \lsi{} shows periodic radio outbursts \citep{gregory99}, whose phase was also reported to vary periodically with the same period as the X-ray peak \citep{gregory99,jaron13}}. More recently, \citet{massi20} also performed a campaign of simultaneous multi-wavelength observations of the system along one single orbit, confirming not only the predicted double-peak light-curve, but also that X-ray dips were coincident with radio and $\gamma$-ray peaks.

There are two main competing scenarios to explain the multi-wavelength observations of \lsi{}, with a special focus on the origin of its non-thermal emission: accretion onto a CO and jet ejection (i.e., a NS or BH microquasar, as first proposed by \citealt{Taylor1982}), or the interaction of a pulsar with the wind of the companion star \citep[first proposed by][]{maraschi81}. 

In the microquasar model \citep[see e.g.,][]{bosch06}, high-energy emission can be produced in jet recollimation shocks that form when the jet crosses the stellar wind of the companion \citep{perucho08,perucho10}, or even in a chain of these type of shocks \citep{miralles22}, that could lead to efficient particle acceleration \citep{rieger07} and synchroton non-thermal emission, inverse Compton and even proton-proton collisions. In this model, the resolved radio structures and the flat radio spectrum shown, for example, by \cite{massi12} and \cite{zim15}, can be interpreted as radio emitting relativistic jets. Several VLBI images \citep{hjellming88,massi04,dhawan06,massi12,wu18} also show that the jet-like morphology changes from one-sided to double-sided, compatible with variable Doppler (de)boosting due to changes in the jet orientation with respect to the line of sight. 
The main drawback of this model is that it has not been possible to confirm any direct proof of accretion in \lsi{}, like a disk black-body component in the energy spectrum or a cutoff power-law spectrum in the high-energy band.

By contrast, \cite{dubus06} and \cite{dhawan06} interpreted the one-sided radio jet of \lsi{} as the cometary tail resulting from the interaction between a pulsar wind and the companion wind, in analogy with the system PSR B1259$-$63 \citep{wang04} that hosts a fast-rotating non-accreting NS with strong magnetic fields. The hypothesis that \lsi{} contains a NS with strong magnetic fields was also proposed by \cite{torres12}, who reported on a Swift-BAT detection of a short burst that resembles those generally labelled as magnetar-like events. More recently, \cite{weng22} found transient radio pulsations of $P \sim296$~ms using observations from the Five-Hundred-meter Aperture Spherical radio telescope. \JLM{However, these pulses seem to be faster than the typical spin period of a magnetar (which usually ranges in the order of several seconds) and were not present in three out of four observations of the source.}

Apart from the source orbital period \JL{($P_{1}$)}, \lsi{} shows other  periodicities. \cite{massi12} first estimated with radio astrometry a precession period of the radio structure of 27-28 days.
Timing analysis of long-term radio flux data from several archives has resulted in the detection of a signal with period $\sim 26.9$~d \citep{massi13, massi16, jaron18}, which is close to the orbital period but still significantly different, and which is in agreement with the previously estimated precession period of the jet \citep{massi12}. 
Analysis of the VLBI astrometry has revealed that the core indeed traces an ellipse with a period of \JL{$P_{2}=26.926\pm0.005$}~d \citep{wu18}. The same precession period has also been detected at X-rays \citep{dai16} and high-energy gamma-rays \citep[][and references therein]{jaron18}.

The simplest explanation is that the observed flux density from a relativistic jet is the product of an intrinsically variable jet and Doppler boosting towards the observer \citep{massi14}. The radio outburst also exhibits a long-term periodic modulation of $P_{\mathrm{long}}=1667\pm8$~days \citep{Gregory2002}, possibly due to the beat of the two close periods, \JL{$P_{1}$} and \JL{$P_{2}$}. Using 6.7 years of data from the Green Bank Interfermoter (GBI), \cite{massi13} suggested that the beating between \JL{$P_{1}$} and \JL{$P_{2}$} also leads to a new apparent periodicity, $P_{\mathrm{av}}=26.70\pm 0.05$ days, which is modulated by $P_{\mathrm{beat}}=1667\pm 393$ days and that is not directly detected in the periodograms (but see also \citealt{massi13} and \citealt{ray97}).

In this paper, we analyse the whole archival RXTE/PCA data of \lsi{}. Our main objective is to present a complete study of the X-ray spectral and fast time variability of the source over the years of exposure, by means of phase-resolved spectral and timing analysis for enhanced counts statistics.

The paper is organized as follows: In Sec. \ref{obsdata} we describe the RXTE/PCA observations that we use in this work and the methodology we follow to split the dataset into three independent sub-intervals. We also describe the phase-resolved analysis and the techniques we employ to produce phase-folded light curves and phase-averaged energy spectra and power density spectra (PDS) for different phase bins. In Sec. \ref{discussion}, we discuss the main results of the analysis and we compare our work with previous results in the X-ray wavelength. In Sec. \ref{conc}, we summarize our results and we draw our main conclusions.

\section{Observations and data analysis}
\label{obsdata}

We analysed all the available \lsi{} X-ray observations performed with the Proportional Counter Array (PCA) instrument \citep{jahoda06} on-board the RXTE satellite. This data set comprises 527 observations covering a broad time interval that extends from 01-03-1996 to 29-12-2011\footnote{After a first inspection of the archive, we discarded some observations because of very short exposures or lack of data. The ObsId of these observations are: 10172-08-01-00, 93100-01-33-00, 95102-01-54-00 and 96102-01-17-00.}, providing a total exposure of over 850 ks. A light-curve of the source (in units of counts/s) including all the observations that we considered is shown in Fig.~\ref{periodogram}~(a) (see Sec.~\ref{lightc} for details on the light-curve extraction). 

\subsection{\JL{RXTE/PCA periodicity}}
\label{observations}

\begin{figure}
	\includegraphics[width=\linewidth]{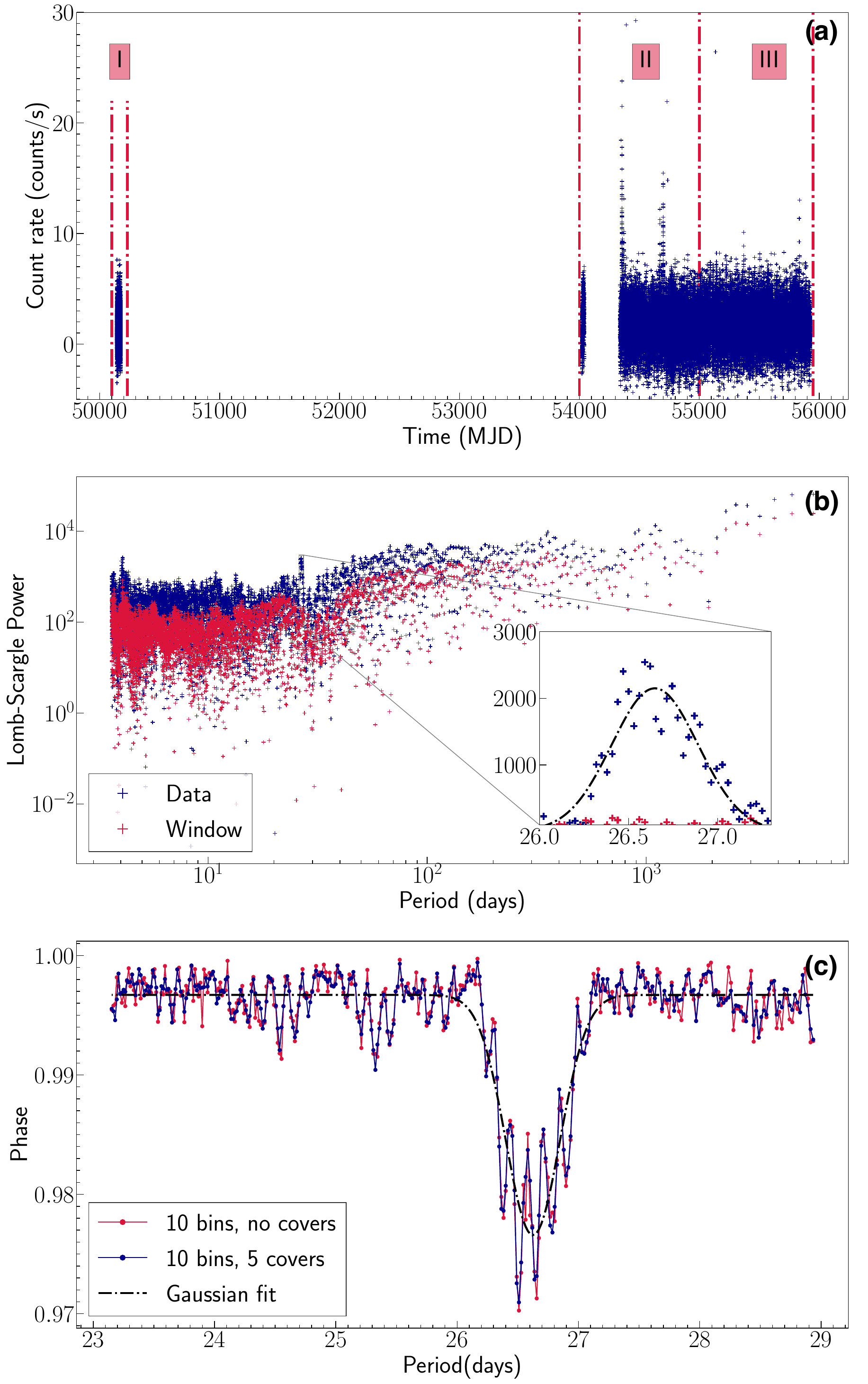}
    \caption{(a) Total light curve with vertical red dash-dot lines showing the start and the end of our three subintervals. (b) Lomb-Scargle periodogram. The LS periodogram shows both the transform of RXTE observations (blue crosses) and the window function (red crosses). The inset plot shows a zoom of the periodic signal, where a Gaussian fit of the peak yields P$=26.65\pm0.28$ days. (c) Phase Dispersion Minimization analysis using 10 phase bins with 5 phase-shifted set of bins (red dot line) and with no phase cover (blue dot line). The peak of the Gaussian fit yields P$=26.63\pm0.26$~days.}
    \label{periodogram}
\end{figure}

In order to confirm if the \JL{periods introduced in Sec.~\ref{sec:introduction}} are also intrinsic periodicities of our X-ray dataset (for the sake of correctness, we should not accept a priori that radio and X-rays show exactly the same orbital modulation), we first produced a Lomb-Scargle (LS) periodogram \citep{Lomb76,scargle89}, which is suitable  for detecting and characterizing periodic signals in unevenly sampled data. Fig.~\ref{periodogram}~(b) shows the periodogram (LS Power vs. period) for the data set (dark blue crosses) and for the window function (red crosses). The latter is computed to determine what features are intrinsic to the data, and what are instead an artefact introduced by the characteristics of the window (i.e. the data sampling). The minimum frequency sampled by the LS periodogram  is $f_{\text{min}}=1/(t_{\text{max}}-t_{\text{min}})=2\times 10^{-9}$~Hz (which corresponds to $P\approx 5781$ days) and the maximum frequency is $f_{\text{max}}=3.18\times 10^{-6}$~Hz (which corresponds to $P\approx 3.6$ days). The frequency resolution is $\Delta f= 4~f_{\text{min}}$ (i.e., we use an oversampling factor of 4). The LS periodogram shows a narrow peak that we fitted with a Gaussian function as shown in the inset plot in Fig.~\ref{periodogram}~(b). The central period of the Gaussian curve is $P = 26.65\pm0.28$ days, where the error corresponds with the Gaussian full width at half maximum (FWHM). 

The LS periodogram is optimized to identify sinusoidal-shaped periodic signals. Essentially, the LS method fits a sinusoidal model to the underlying data at each frequency, with a larger power reflecting a better fit. However, for some signals, the assumption of stationary sinusoidal models could lead to inaccurate results, besides the limitation in frequency resolution imposed by the data time coverage. Thus, to confirm the period found in the LS periodogram of Fig.~\ref{periodogram}~(b), we repeated the analysis using two complementary statistical methods: (1) the Phase Dispersion Minimization (PDM) method\footnote{\url{https://pyastronomy.readthedocs.io/en/latest/pyTimingDoc/pyPDMDoc/pdm.html}} \citep[][Fig.~\ref{periodogram}~c]{pdm78}  and (2) amplitude maximization with a sinusoidal fit, which is not shown in Fig. \ref{periodogram}. The first technique finds periodic variations by minimizing the dispersion of the folded dataset and it is commonly used to analyze time series with gaps, non-sinusoidal variations, poor time coverage or when the Fourier techniques lead to wrong solutions. For this particular case, we used 10 phase bins with 5 phase-shifted sets of bins (dark blue line) and with no phase covers (red line), for comparison. In both cases, the results obtained are very similar. A Gaussian fit over the phase-shifted data yields $P=26.63\pm0.26$~days, in good agreement with the LS period. In the second complementary method, we directly searched for the best sinusoidal fit parameters of the phase-folded light curve modulation in the period range $26.0-28.0$~days, using 400 fit trials. Considering the following sinusoid, $y=a_0+a_1\sin{(a_2t+a_3)}$, the period that maximizes $|a_1/a_0|$ represents the best possible data modulation. In this case, the peak of the Gaussian fit yields a central period P$=26.62\pm0.30$~days, in good agreement with our two previous estimations.

\subsection{Data analysis}


\JL{We aim to track any possible variation of the spectral and timing features of the source over 15 years, and hence  we phase-folded the observations with a given X-ray period to increase the signal-to-noise (S/N) ratio of each phase bin. Assuming no relevant variations occur at a particular phase, this method allows us to obtain averaged phase resolved energy spectra and PDS with enhanced statistics.}

\JL{We divided our data set into three intervals, intended to maintain a good S/N ratio while we avoid the smearing of any possible long-term variability. We considered the following subsets: Interval I, that includes all the March 1996 observations (identified with proposal number 10172), and Intervals II and III, each including about half of the remaining observations (covering more than five years in the time range from 13-10-2006 to 29-12-2011). The number of observations in each interval, the time covered and the total exposure are listed in Tab.~ \ref{tab:1}.}

\JL{The phase associated to any given periodicity $P$ is given by $ \phi_{\text{}}=\left(T_{\rm REF}+T-T_0\right)/P-\text{int}\left[\left(T_{\rm REF}+T-T_0\right)/P\right]$, where $T_0 = 43366.275$ MJD is the time of the first  radio detection of the source, $T_{\rm REF} = 49353.00069657407$ MJD is the RXTE reference epoch and $T$ is the spacecraft clock reading (including clock corrections).}

\JL{Data filtering was performed with the standard criteria typically applied to RXTE/PCA\footnote{\url{https://heasarc.gsfc.nasa.gov/docs/xte/recipes/cook\_book.html}}, while data analysis was carried out using HEASoft 6.28. For light-curve extraction (Sec. \ref{lightc}) and spectral fitting (Sec. \ref{xspec}), only data from the Proportional Counter Unit-2 (PCU-2) was  used for the analysis. This is commonly done as the PCU-2 is the only detector that always kept on along the entire duration of the RXTE mission.}

\begin{table*}
	\centering
	\caption{Main parameters of the three subsets of observations considered in this work. The number of observations of the 1996 campaign is much lower than Interval II and Interval III, but the total exposure of the subset is comparable within a factor.}
	\label{tab:1}
	\begin{tabular}{lcccr} 
		\hline
		Subset & Number of observations & Start date (ObsId) & End date (ObsId) & Exposure\\
		\hline
		Interval I & 12 & 01-03-96 (10172-01-01-00) &30-03-96 (10172-11-01-00) & 106 ks\\
		Interval II & 255 & 13-10-06 (92418-01-01-00)  &11-06-09 (94102-01-30-00)& 397 ks\\
		Interval III & 254 & 14-06-09 (94102-01-31-00) &29-12-11 (96102-01-06-10) & 360 ks\\
		\hline
	\end{tabular}
\end{table*}

\subsubsection{Light curves}
\label{lightc}

\begin{figure}

	\includegraphics[width=\columnwidth]{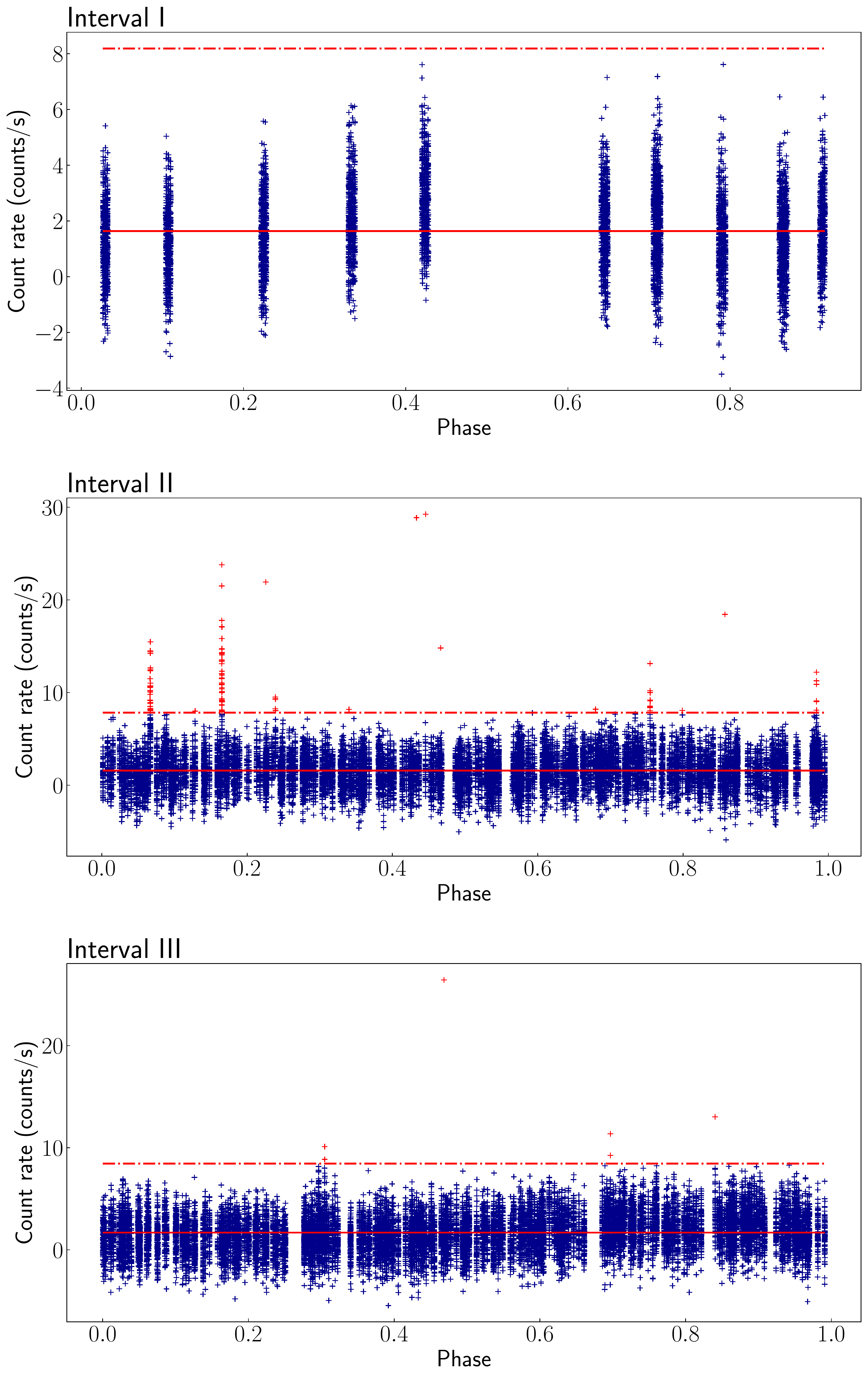}
    \caption{Unbinned phase-folded lightcurve for Interval I (top), Interval II (middle) and Interval III (bottom). The mean count rate of each interval is represented with a red solid line. Detections above the red dashed-dot line (which is placed at five times the average count rate) are not considered for light curve rebinning. These points could be either real events associated to the source flaring activity (e.g., the big flares of Interval II) or non-physical instrumental features that do not show any hint of phase dependency.}
    \label{flares}
\end{figure}

For each of the three sub-intervals in Tab.~\ref{tab:1}, we produced a phase-folded light curve using Standard 2 type data, which is characterized by low time resolution (16s) and moderate energy  resolution ($<$ 18\% at 6 keV, with 129 energy channels covering the nominal energy range 2-120 keV). We selected canonical Good Time Intervals (GTI) by choosing the times when the source elevation was >10$^{\circ}$ and the pointing offset was <0.02$^{\circ}$. We estimated the background using \textsc{pcabackest} v3.12a and the most recent background file available on the \textsc{heasarc} website for faint sources \footnote{pca\_bkgd\_cmfaintl7\_eMv20151128.mdl}. Source and background light curves were then extracted using the \textsc{ftool} \citep{black95} software package utility \textsc{saextrct} by selecting channels in the range~4-128 (see the discussion in Sec. \ref{timing} about low energy channels). We estimated the background-subtracted light curve for each observation using the \textsc{ftool} routine \textsc{lcmath}. Barycentric corrections were applied using the routine \textsc{barycorr} and the ephemeris file \textsc{JPLEPH.430}, which contains the most up-to-date solutions as of the writing of this paper.
To fold the light curve in phase, we calculated the phase for each point in the curve using the period we measured in Sec. \ref{observations}. For the sake of consistency, and given the uncertainties of our statistical methods, hereinafter we consider P$=26.6$~days. \JLM{We have checked that no significant differences in the phase modulation appears when considering instead the orbital period, P$_1$}. This procedure was then repeated for all the observations in the subset to produce one single unbinned phase-folded light curve for the entire duration of the interval, as shown in the three panels of Fig.~\ref{flares}. Then, we divided the data into 10 phase bins of width 0.1 (i.e., $2.66$~days/bin), and we estimated the mean count rate in each of them. Outlier points (i.e., detections where the count rate exceeded more than five times the mean of the light curve) were removed before rebinning the light curve in phase, since these might affect the source phase modulation (see, for example, Fig. 2 in \citealt{Li11}). These outlying points correspond to short flares that were previously identified by other authors \citep[see e.g.,][]{smith09,Li11}. Fig.~\ref{flares} shows the location of these flares, besides other random outliers (red points), for each of the three sub-intervals that we analysed in this paper. Apart from the big flares of Interval II grouped in the 0.0-0.3 and 0.7-1.0 phase bins, the light curves also show a few, likely instrumental, smaller deviations that do not appear to be preferentially observed at any particular phase.

\subsubsection{Power density spectra}
\label{timing}

We used RXTE Good Xenon data mode, which are science event format files characterized by very high time ($\sim$ 0.95 $\mu$s) and energy resolution (256-channel pass band with energy resolution $<$ 18\% at 6 keV). The Good Xenon configurations use two Event Analyzers (EA) simultaneously to provide detailed spectral and temporal information about every event that survives background rejection. Each EA creates matched pairs of files that we merged with the Perl script \textsc{make\_se}, prioritizing data with a readout time of 16s over 2s, depending on availability. We phase-folded all the observations in the data set (using the X-ray intrinsic period P~$=26.6$~days), obtaining ten different phase bins as described in the previous section. For each of these bins, we produced a PDS using the custom GHATS package\footnote{See  http://www.brera.inaf.it/utenti/belloni/GHATS\_ Package/Home.html.}. PDS were extracted only for energy channels in the range 8-255 \footnote{https://heasarc.gsfc.nasa.gov/docs/xte/e-c\_table.html}($<3.7$~keV). We avoided channels 1-7 because we found that its inclusion in the timing analysis produced artefacts in the PDS which we could not mitigate (i.e., PDS were distorted with undesirable drops near the Nyquist frequency). The effective area of the PCA is small at these energies, therefore the exclusion of events in these bands does not change significantly the amount of photons collected, especially after phase-folding.  On the other hand, we also removed sharp drop outs from the data (that usually appear at the beggining or at the end of an observation), whose nature is instrumental and not related with the source. If present, the existence of instrumental drops can introduce artificial steep features in the low-frequency noise. This task was performed by a specific selection of GTIs based on three fundamental steps: (1) first, we selected RXTE standard GTIs using \textsc{maketime} based on housekeeping information, by choosing the times when the source elevation was >10$^{\circ}$ and the pointing offset was <0.02$^{\circ}$.
(2) Secondly, we used a bi-weight algorithm\footnote{The bi-weight algorithm calculates the center and dispersion of a distribution using bisquare weighting.} to detect and handle outliers, for which we considered a threshold of $\pm 3.5\sigma_{\rm bi}$, where $\sigma_{\rm bi}$ is the bi-weight standard deviation. When an outlier is detected, \JL{we remove an interval of $\pm 16$~s. around the outlier point}. Intervals with a time length lower than the previous \JL{threshold of $16$~s.} were automatically discarded.
(3) Finally, we used the \textsc{ftools} utility \textsc{ftmgtime} to merge by intersection the GTI files obtained in the two previous steps. 

All event files and GTIs were barycenter corrected with the \textsc{ftool} utility \textsc{fxbary}, but no deadtime corrections nor background subtractions were performed before creating the PDS due to the very low count rate. Using Fast Fourier Transform (FFT) techniques, we rebinned the data in time to obtain a Nyquist frequency of 4096 Hz and we produced PDS for continuous 256-s-long data segments (leading to a frequency resolution of $1/256$ s $\approx 0.004$ Hz). For each phase bin, we averaged all the PDS normalized according to \cite{1983ApJ...266..160L} and subtracted the Poisson noise spectrum following \cite{1995ApJ...449..930Z}. Finally, we estimated the PDS root mean squared deviation (rms) in a low frequency ($0.006-0.1$ Hz) and a high frequency ($0.1-512.0$ Hz) band as a function of phase, in order to investigate the variability of the emission. Background count rate was estimated from the background light curve (see the details on Sec. \ref{lightc}); for those phase bins which contained only one observation, the mean background was calculated as the mean count rate of the background light curve. For bins including more than one observation, we first calculated the mean of each individual observation and then, after checking that no large differences existed in the average background rates from individual observations, we computed the average and standard deviation of the bin. 

The objective of this analysis is to search for timing features in the data (i.e., broad-band red noise components, quasi-periodic oscillations, etc), for which we aim to maximize the S/N. Therefore, for this section, we only consider -and thus we only show in the paper- the PDS calculated with the whole set of observations. The rms, however, is splitted in the three subintervals of Table \ref{tab:1}.


\subsubsection{Spectral fitting}
\label{xspec}

We calculated source and background spectra for each individual observation using the \textsc{ftools} utility \textsc{saextrct} using Standard 2 data, applying the same GTIs described in Section \ref{timing}. Dead time corrections were applied using the standard RXTE procedures\footnote{https://heasarc.gsfc.nasa.gov/docs/xte/rcrsp} v11.7.1 and we averaged  the source and background spectra in each of the ten phase bins using the \textsc{fortran} wrapper \textsc{addspec} v1.4.0 to obtain one source and one background averaged spectrum per bin. Then, we fitted the background-subtracted spectra between 4 and 30 keV in \textsc{xspec} v12.11.1 \citep{xspec} using three model components: the interstellar photoelectric absorption (\textsc{tbabs}), one power-law (\textsc{power}) to fit the source signal at low energies and a second power-law (\textsc{power}) to fit the galactic ridge emission, which is present in all energy spectra dominating the emission above $\sim$20 keV \citep{revnivtsev07}. The absorption coefficient $N_H=7.8\times 10^{21}$ cm$^{-2}$ and the photon index of the second power-law $\Gamma=0.0$ were frozen before fitting the model to the spectra. 
As in the previous section, we only show the averaged energy spectra obtained using the whole set of observations, although the photon index is also calculated in the three sub-intervals defined in Tab.~\ref{tab:1}.

\subsection{Results}

\begin{figure}

	\includegraphics[width=\columnwidth]{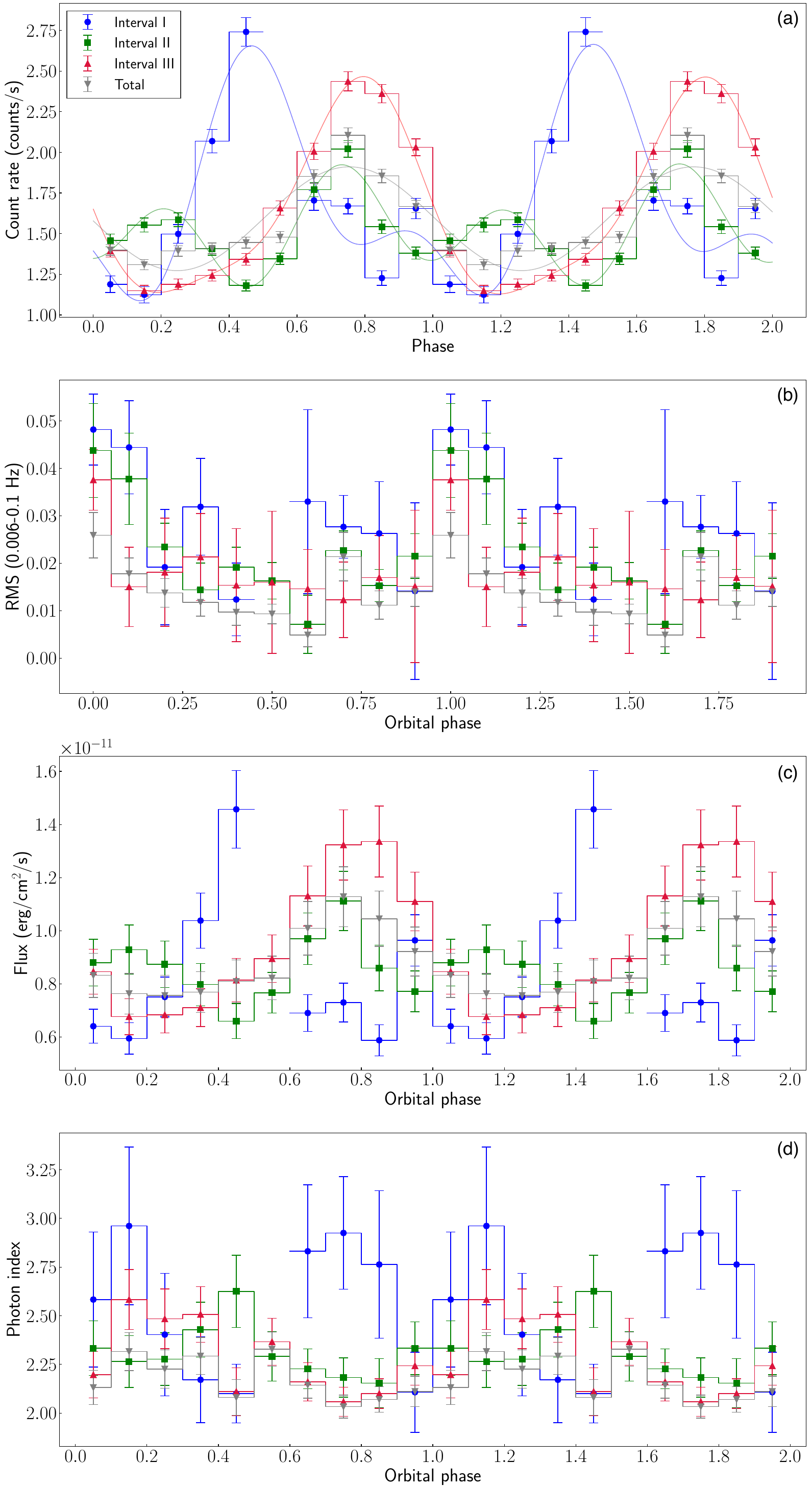}
    \caption{RXTE/PCA data vs. phase: (a) light curve count rate (counts/s) \JL{over-plotted with the best sinusoidal fitting shown in Tab.~\ref{fit}}, (b) low frequency rms (0.006-0.01 Hz). In Interval III, the rms is multiplied by a correcting factor of 5 (c) 4-10 keV flux (erg/cm$^2$/s), (d) photon index. There is a gap in Interval I because of lack of data in the phase bin $0.5-0.6$. }
    \label{results}
    
\end{figure}

The main results of our data analysis are shown in Fig. \ref{results} for Interval I (blue circles), Interval II (green squares), Interval III (red up-triangles), and the total dataset (grey upside-down triangles), which is also shown for the sake of comparison. 

Fig.~\ref{results}~(a) shows the light curve in units of counts/s as a function of phase (with respect to the measured period), \JL{where error bars include both the Poisson error and a $2\%$ systematic uncertainty to account for the low number statistics}. We note that Interval I shows a gap at phase $0.5-0.6$ because there is no data in the archive for this particular bin. All light curves show a clear modulation which is consistent with the orbital periodicity, with the count rate reaching its minimum in the bin $0.1-0.2$ for Interval I \JL{($1.124\pm0.045$~counts/s)} and Interval III \JL{($1.149\pm0.023$~counts/s)}, and in the bin $0.4-0.5$ for Interval II \JL{($1.182\pm0.025$~counts/s)}. Interval I shows a maximum in the bin $0.4-0.5$ \JL{($2.741\pm 0.068$ counts/s)}, which is shifted in phase to bin $0.7-0.8$ for Interval II \JL{($2.021\pm0.031$~counts/s)} and Interval III \JL{($2.437\pm0.032$~counts/s)}. \JL{For each of these intervals, we fitted a constant and a single or double sinusoidal function, where the best fit parameters are given in Tab.~\ref{fit}. The two-wave sinusoidal function aims to test statistically the accretion models that predicts a two-peak light-curve, which in this case is more apparent for Interval I and Interval II. In all three cases, statistics improve significantly by fitting the light-curve with the sum of two sinusoids, but the goodness of fit is only statistically acceptable for the double sinusoidal function in Interval III. 
}

\begin{table*}
    \caption{Best fit parameters of the sinusoid functions used in Fig. \ref{results} (a), where $H$ is the vertical shift, $|a|$ is the wave amplitude, $\omega$ is the angular frequency and $\phi$ is the phase. Sub-indices 1,2 refer to the \JL{single and double} sinusoids used in the analysis, \JL{
    respectively. Last column shows the $\chi_{\nu}^2$ for the constant fit}.}
    \label{fit}    
    \footnotesize{}
    
    \begin{tabulary}{\textwidth}{@{}LCCCCCCCCC@{}}
        \toprule 
        & \multicolumn{9}{c}{{\footnotesize{}Fit parameters}}\tabularnewline   
        \cmidrule{2-10}
        {\bf Sinusoids} & {$H$} & {$a_1$} & {$w_1$} & {$\phi_1$} & {$a_2$} & {$w_2$} & {$\phi_2$} & {$\chi_{\nu}^2$ (dof)}& {$\chi_{\nu}^2$ (dof) -const.} \tabularnewline
         
        \midrule
        {Interval I (1)} & {$1.74$} & {$-0.57$} & {$6.52$} & {$1.37$} & - & - & - & {$21.55 (16)$}& {$50.97 (19)$}\tabularnewline

        {Interval I (2)} & {$1.74$} & {$-0.60$} & {$6.29$} & {$1.56$} & {$-0.33$} & {$12.43$} & {$-0.88$} & {$10.10 (13)$}& {$-$}\tabularnewline

        {Interval II (1)} & {$1.52$} & {$0.17$} & {$6.86$} & {$2.21$} & - & - & - & {$26.85 (16)$}& {$29.95 (19)$} \tabularnewline         
        
        {Interval II (2)} & {$1.53$} & {$0.26$} & {$12.62$} & {$-7.52$} & {$0.16$} & {$6.39$} & {$-3.57$} & {$3.05 (13)$}& {$-$}\tabularnewline  
        
        {Interval III (1)} & {$1.68$} & {$0.63$} & {$6.13$} & {$3.12$} & - & - & - & {$9.80 (16)$}& {$125.96 (19)$} \tabularnewline  

        {Interval III (2)} & {$1.68$} & {$0.64$} & {$6.25$} & {$3.00$} & {$0.16$} & {$12.44$} & {$-8.58$} & {$0.87 (13)$}& {$-$} \tabularnewline  
               
        \bottomrule
    \end{tabulary}

\end{table*}

\begin{figure}
 	\includegraphics[width=\columnwidth]{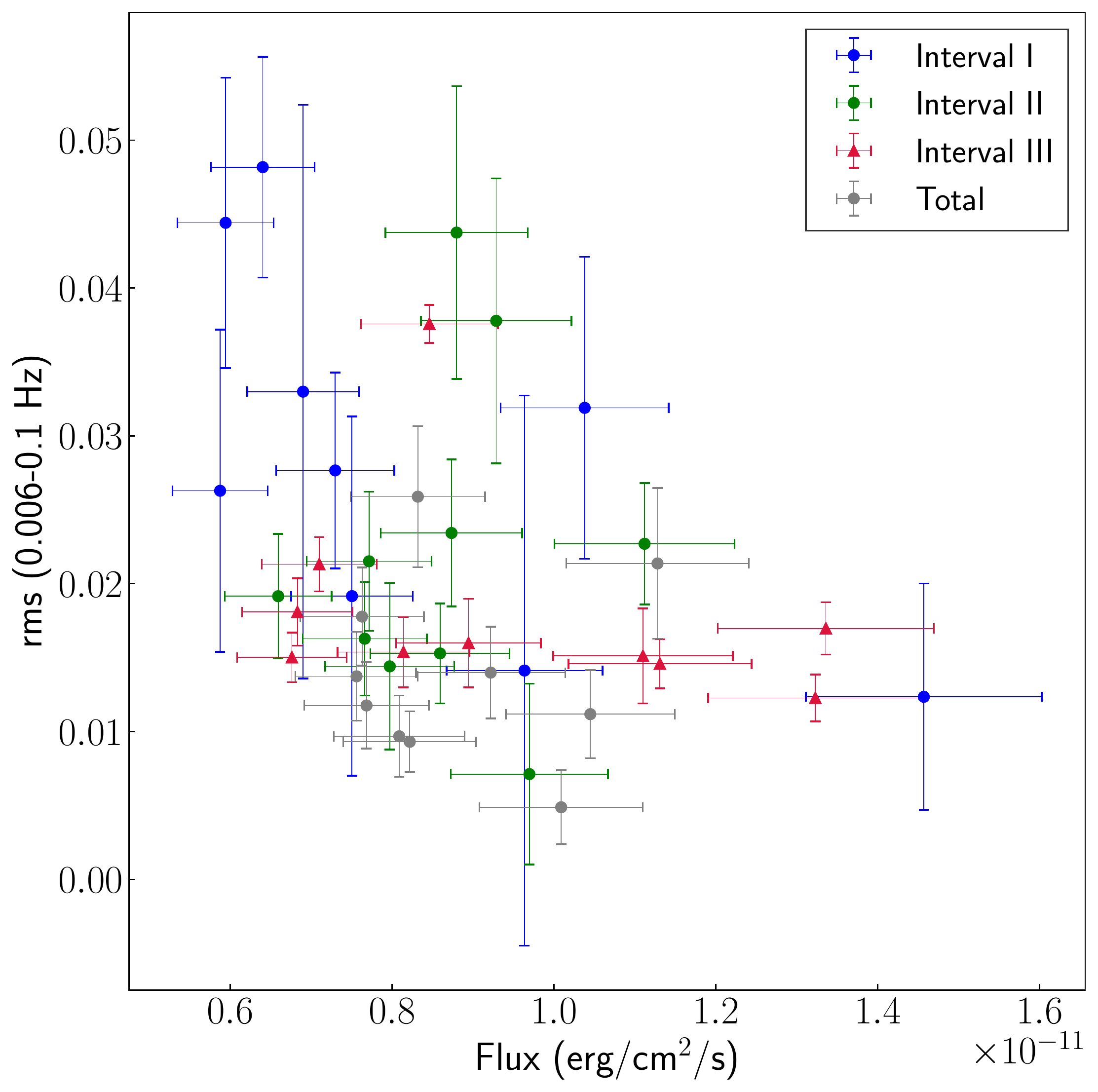}
	\includegraphics[width=\linewidth]{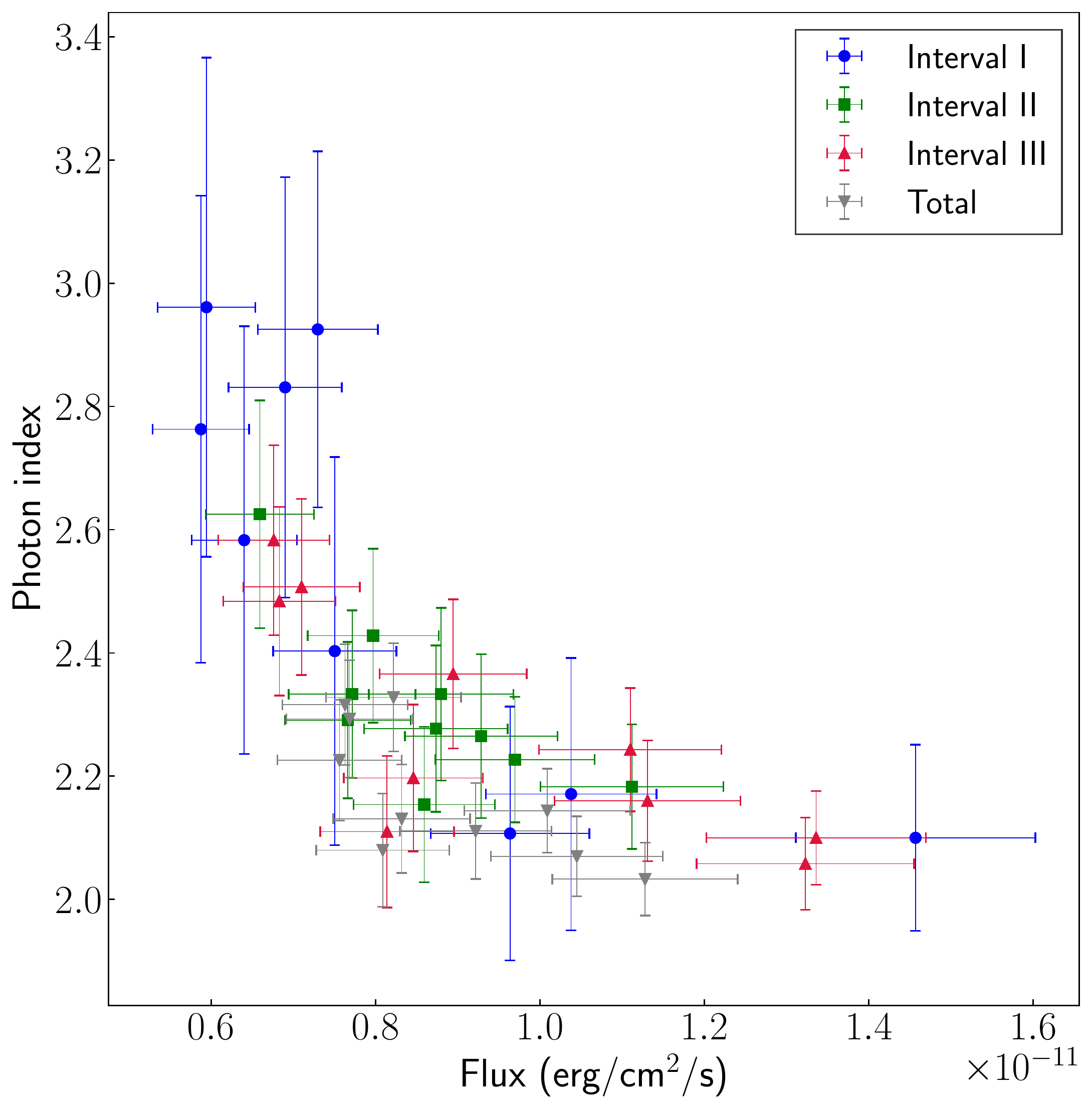}
    \caption{\JL{Top: Fractional rms in $0.006-0.1$~Hz vs. 4-10 keV flux. Bottom: Photon index  vs. 4-10 keV flux, where data shows a negative strong correlation.}}
    \label{vs}
\end{figure}

Fig.~\ref{results}~(b) shows the rms measured for every phase bin as described in Sec. \ref{timing}. Although we calculated the rms in two frequency ranges ($0.006-0.1$ Hz and $0.1-521$ Hz), we only report the low-frequency segment since the high-frequency rms is massively dominated by the instrumental noise of the detectors.
There are hints for rms phase dependency in the three intervals, where the rms maximum always appears in the bin $0.0-0.1$. We note, however, that the overall rms of Interval III is considerably lower than in the other two intervals. Indeed, in Fig.~\ref{results}~(b) Interval III rms is displayed multiplied by a factor 5 to facilitate the inspection of the curve as compared with the other two intervals. We could not identify a clear reason why the variability in this interval is overall significantly lower than in the others, although we speculate that this behaviour could be attributed to the degradation of the instruments over time.  Overall, the bumps in the rms seems to occur \JL{near the minimum of the X-ray flux (or light curve count rate), although the large uncertainties prevent us to firmly establish the presence of any statistically significant correlation with phase. A constant fit gives in this case $\chi^2_{\nu}=1.62(19)$ for Interval I, $\chi^2_{\nu}=2.35(19)$ for Interval II and $\chi^2_{\nu}=2.49(19)$ for Interval III. This means that, at least for Interval I (where error bars are larger), the low-frequency rms is statistically compatible with a constant, but Interval II and Interval III show weak variability. In order to investigate the goodness of the correlation between the X-ray flux and the rms, we have estimated the Spearman’s rank correlation coefficient \citep{curran14}, using data from all intervals together. The test yields $\rho=-0.4$, which indicates a weak negative correlation (see also Fig. \ref{vs}, top panel), but it can be increased up to $\rho=-0.6$ considering only data from Interval I (which is however affected by larger uncertainties).}

The X-ray flux (in units of erg/cm$^2$/s), which has been extracted in the energy range $4.0-10.0$ keV assuming an error of 10\% to be conservative with the response of the instrument, is shown in Fig.~\ref{results}~(c). As expected, the modulation of the flux is consistent with the count rate modulation seen in Fig.~\ref{results}~(a) for all intervals. The flux shows a minimum in the bin $0.1-0.2$ ($0.59\pm 0.06\times 10^{-11}$ erg/cm$^2$/s) and $0.8-0.9$ ($0.58\pm 0.06\times 10^{-11}$ erg/cm$^2$/s) for Interval I, in the bin $0.4-0.5$ for Interval II ($0.66\pm 0.07\times 10^{-11}$ erg/cm$^2$/s), and in the bin $0.1-0.2$ for Interval III ($0.68\pm 0.07\times 10^{-11}$ erg/cm$^2$/s). The maximum flux occurs at $0.4-0.5$ for Interval I ($1.46\pm 0.015\times 10^{-11}$ erg/cm$^2$/s), at $0.7-0.8$ for Interval II ($1.11\pm 0.11\times 10^{-11}$ erg/cm$^2$/s), and at $0.8-0.9$ for Interval III ($1.34\pm 0.13\times 10^{-11}$ erg/cm$^2$/s). As in Fig.~\ref{results}~(a), there is a second peak at bin \JL{$0.9-1.0$ for Interval I ($0.96\pm 0.10\times 10^{-11}$ erg/cm$^2$/s)} and at bin $0.1-0.2$ for Interval II ($0.93\pm 0.09\times 10^{-11}$ erg/cm$^2$/s), which provides more significance to the two-peak light-curve modulation.

Fig.~\ref{results}~(d) shows the evolution of the source photon index as a function of phase. The photon index is anti-correlated with respect to both the count rate and the X-ray flux, being maximum near the first and last  bins for our three intervals. This anti-correlation becomes evident when the photon index is plotted as a function of the flux, as shown in the bottom panel of Fig. \ref{vs}. In order to investigate the goodness of this correlation, we have also estimated the Spearman’s rank correlation coefficient, using data from all intervals. The test yields $\rho=-0.87$, which indicates in this case a strong anti-correlation. The linear fits to the data have slopes $-1\pm0.3$ ($\chi^2$/dof=$6.21/7$) for Interval I, $-0.81\pm0.3$ ($\chi^2$/dof=$3.85/8$) for Interval II and $-0.59\pm0.2$ ($\chi^2$/dof=$7.85/8$) for Interval III, such that the slope experiences a smooth flattening from the former to the latter interval.

\begin{figure*}

	\includegraphics[width=\textwidth]{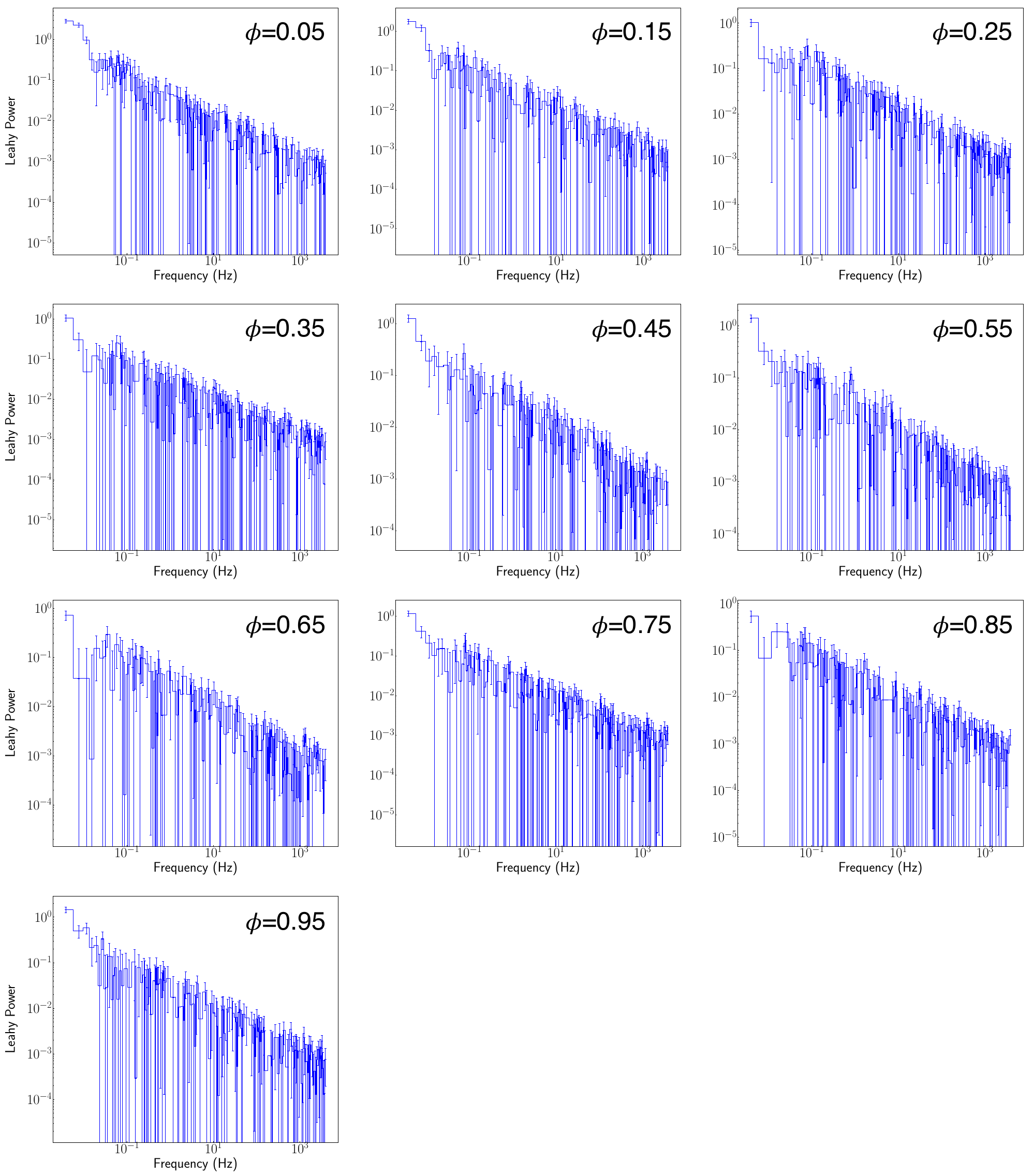}
    \caption{Leahy normalized poisson noise-subtracted PDS averaged on ten phase bins using the entire X-ray dataset. The signal from the source is significant only below 0.1 Hz, above which all PDS are dominated by the instrumental noise.}
    \label{POW}
    
\end{figure*}

In Fig.~\ref{POW}, we show a collection of PDS for the ten phase bins calculated with the entire data set. Save for low-frequency red noise below $0.1$ Hz, there are no statistically significant features in any of the PDS we extracted (i.e., other types of broad-band noise or QPOs). There is only a weak narrow feature that looks more prominent in the bin $\phi=0.7-0.8$, at $f\sim0.08$~Hz. However, a Lorentzian fit of the feature gives no statistical significance ($\sigma\sim2$).

\begin{figure*}

	\includegraphics[width=\textwidth]{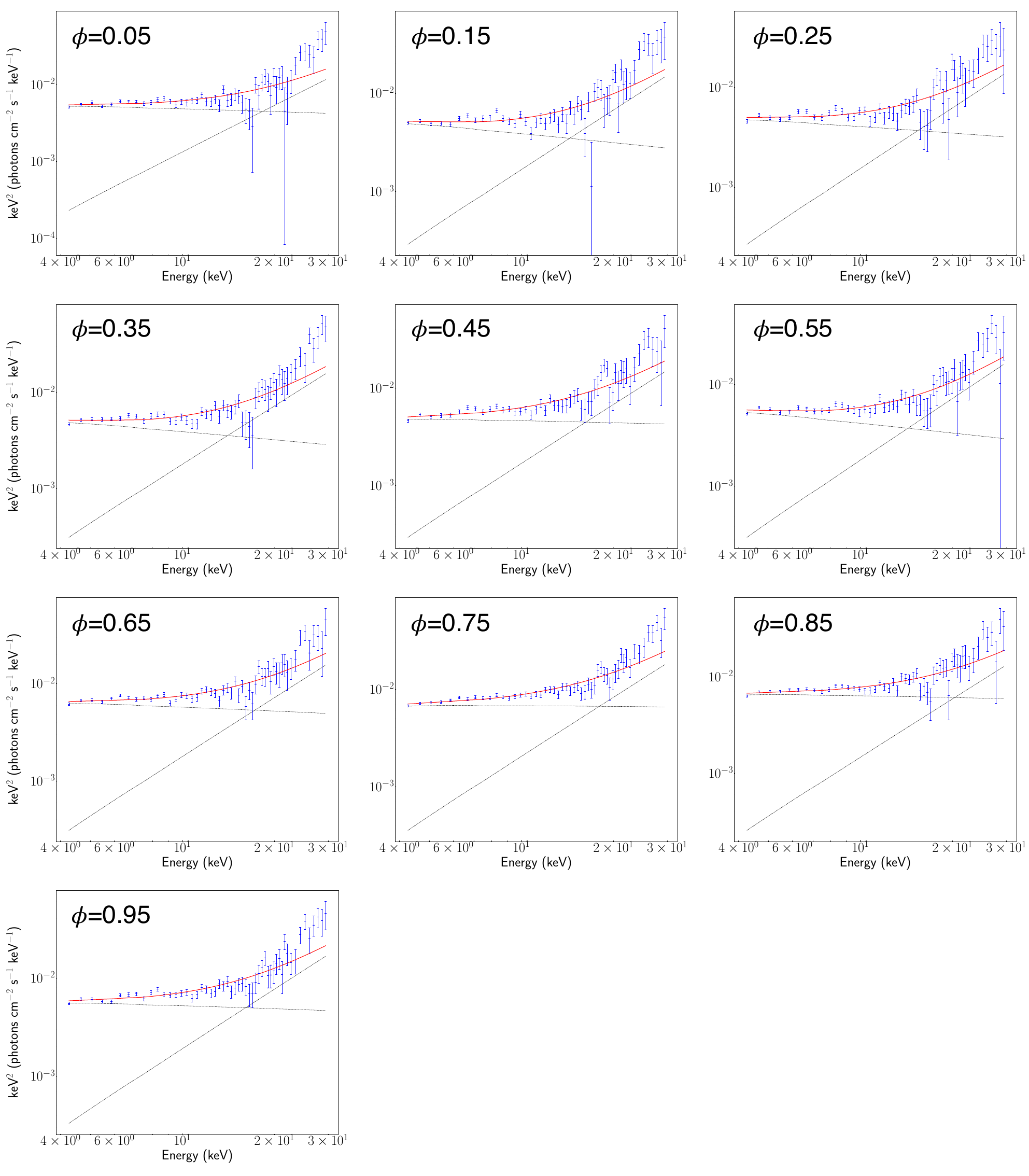}
    \caption{Energy spectra averaged on ten phase bins using the entire X-ray dataset. The statistical model (solid red line) results from the combination of three model components: the interstellar photoelectric absorption and two power-laws (dashed black lines).}
    \label{SED}
    
\end{figure*}

A collection of energy spectra, also for the entire set of observations, is given in Fig. \ref{SED}. In the plots, we have represented the two power-law components (black dashed line) and the combined statistical model (red solid line). In general, energy spectra pertained to different phase bins do not show significant variations.

\section{Discussion}
\label{discussion}

The gamma-ray binary \lsi{} has been extensively studied at different wavelengths since its discovery more than sixty years ago. However, the nature of the compact object and the reasons for its peculiar behavior, unique among other galactic binaries, are still debated. 

In this paper, we have analysed all the available RXTE/PCA X-ray data of \lsi{}, i.e. over 500 individual observations taken between 1996 and 2011, in order to investigate the spectral and fast time variability of the binary. Since \lsi{} is a faint source that shows a clear phase modulation, we performed a phase-resolved analysis folding on its intrinsic period to calculate averaged energy spectra and PDS with enhanced S/N. For the sake of full consistency, we measured the periodicity in the RXTE/PCA data in order to avoid biases by adopting previous estimates of the period obtained in other energy bands, or by other instruments in X-rays. Since significant time variations in the X-ray modulation have been reported in previous studies \citep[see e.g.,][]{Torres10,Li11}, our analysis was performed considering three different intervals (see Tab.~\ref{tab:1}). The number of divisions, and thus the time length of each interval, aims to blur any possible trend in the data, but keeping an acceptable S/N for the analysis, as explained in Sec.~\ref{obsdata}.

The results of the three timing methods described in Sec.~\ref{observations} are shown in Tab.~\ref{tab:periods}, together with a list of  periodicities found in the literature, at different wavelengths and within the time interval 26-27 days. For each value, we report the energy band, instrument and statistical method used to determine the period. Even though the results of \cite{Torres10} using RXTE/PCA data (i.e., P$=26.68\pm0.58$~days) and \cite{ray97} based on GBI radio observations (i.e., P$=26.69\pm0.02$~days) are the closest to our central period, all the results reported in Tab.~\ref{tab:periods} are compatible with our measurements within 1$\sigma$, so we cannot exclude  that our period is respectively higher or lower than the well-defined orbital \citep{Gregory2002} and precession periods \citep{wu18}. Nevertheless, we note that despite the high frequency resolution of our dataset, the LS periodogram does not show a double peak distribution as found in radio \citep[see e.g.,][]{massi13} \JLM{or in hard X-rays from Swift/BAT survey data \citep{dai16}}.

\begin{table*}
	\centering
	\caption{Summary of  \lsi{} periodicities found in the literature in the range 26-27 days for different intruments and energy wave-lengths. For each result, we show the statistical method by which the periodicity was calculated.}
	\label{tab:periods}
	\begin{tabular}{lccccc} 
		\hline
		Reference & Period (days) & Error & Energy band & Instrument &Statistical method \\
		\hline
         This work    & \bf 26.65 & \bf $\pm$0.28 & \bf X-ray & \bf RXTE/PCA & \bf LS Periodogram \\
         & \bf 26.63 & \bf $\pm$0.26 & \bf X-ray & \bf RXTE/PCA & \bf PDM \\
         & \bf 26.62 & \bf $\pm$0.30 & \bf X-ray & \bf RXTE/PCA & \bf Sinusoidal fit \\
        \hline
        \cite{jaron18} & 26.45 & $\pm$0.05 & $\gamma$-ray & Fermi-LAT& LS Periodogram \\    
         & 26.99 & $\pm$0.05 & $\gamma$-ray & Fermi-LAT& LS Periodogram \\  
        \cite{leahy01} & 26.42 & $\pm$0.05 & X-ray & RXTE/ASM & Epoch folding \\
        \cite{Torres10} & 26.68 & $\pm$0.58 & X-ray & RXTE/PCA & Power spectrum \\
        \cite{dai16} & 26.47 & $\pm$0.10 & X-ray & Swift/BAT & LS Periodogram \\
         & 26.93 & $\pm$0.10 & X-ray & Swift/BAT & LS Periodogram \\
        \cite{ray97} & 26.69 & $\pm$0.02 & Radio & GBI & Linear fitting  \\
        \cite{Gregory2002} & 26.496 & $\pm$0.0028 & Radio & GBI & Gregory-Loredo Bayesian  \\
        \cite{massi13} & 26.70 & $\pm$0.05 & Radio & GBI & (Aparent)  \\
        \cite{massi16} & 26.496 & $\pm$0.013 & Radio & GBI & LS Periodogram  \\
         & 26.935 & $\pm$0.013 & Radio & GBI & LS Periodogram  \\
        \cite{wu18} & 26.926 & $\pm$0.005 & Radio & VLBA & Pattern alignment  \\
        \cite{zamanov13}& 26.502 & $\pm$0.007 & Optical H$\alpha$ & RCC/Coude & PDM/CLEAN \\
		\hline
	\end{tabular}
\end{table*}

One interesting outcome of our analysis, shown in Fig. \ref{results}, is that the light curve peak is phase shifted ($\Delta\Phi\sim 0.3$) from Interval I to Interval II/Interval III. As expected, the same behaviour is seen both in the light-curve and in the flux. This is consistent with the fact that the X-ray light-curve modulation changes with time, as previously reported by other authors \citep{Torres10,Li11}.  \JLM{Using radio data, \cite{Gregory2002} also reported a super-orbital period in the radio band, P$_{\mathrm{long}}=1667\pm8$~days \citep[see also][for a discussion of this long-term modulation across the electromagnetic spectrum]{jaron21}. \cite{Li11} could not detect such modulation in the system flux history, but \cite{Li12} found evidence for the existence of this period using the longest RXTE/PCA continuous monitoring of \lsi{}. Thus, since we do not apply any specific correction, it is plausible that the underlying long-term periodicity -if really present in the X-ray data- phase-shifts the peak of the light curve from Interval I to Interval II, inasmuch as there is a time gap of 10 years between both intervals. Although the study of the super-orbital period is beyond the scope of the paper, the LS periodogram does not show any signal at that period, likely because our dataset is not long enough to properly sample such a long-term periodicity \JL{using Fourier techniques}. Moreover, we have checked that no significant super-orbital phase modulation appears when folding the light-curve with this long period using the same methodology  described in Sec.~\ref{lightc}}

Regarding phase modulation, theory and numerical modelling predicts two main peaks in the accretion rate of \lsi{} \citep{bosch06,romero07}, each of which should be followed (assuming the microquasar model) by an ejection of particles in the form of a jet that will emit non-thermal synchrotron radiation. \JL{According to the statistical models shown in Tab.~ \ref{fit}, all intervals are better approximated by a sum of two sinusoids. This is especially relevant for Interval II (where the amplitude of the main peak at phase $\theta=0.7$ is lower than in the other two intervals), while in Interval I the secondary peak at $\theta=0.95$ is narrower. In this case, the relevance of the secondary peak is supported by a flux local maximum (Fig.~\ref{results}, c) together with a deep of the photon index (Fig.~\ref{results}, d). Moreover, the phase shift between the primary and secondary peak is similar for Interval I and for Interval II, $\sim 0.5$}.  In the X-ray energy range, the light-curve displayed in Fig. 2 of \cite{smith09} using RXTE/PCA data (taken every other day between 2007 August 28 and 2008 February 2, so it is enclosed in our Interval II) shows a similar phase modulation and a similar ratio between the amplitude of the two accretion peaks. Nevertheless, the authors concluded that there was no statistically strong detection of modulation of the flux with the orbital phase. Later, \cite{Torres10} showed that the two-peak structure evolves into a more clearly visible, single-peaked light curve analysing individual six month periods that partly covered our Intervals II and III. This is also consistent with the fact that the \JL{secondary peak is only marginally detected} in the light-curve of Interval III. \cite{massi20} also found the two-peak light curve analysing Swift-XRT, XMM-Newton and NuSTAR data taken on July 2017, showing that radio and $\gamma-$ ray peaks are coincident with X-ray dips as expected for emitting ejections depleting the X-ray emitting flow. Therefore, RXTE/PCA observations confirm the two-peak light-curve predicted by previous accretion models for \lsi{}. However, our results suggest that either this behaviour is a transient feature of the source, or the amplitude of the secondary peak varies with time.

We also confirm the findings of \cite{Li11}, who reported that the anti-correlation between the flux and the spectral index is an orbit-associated effect, and that this correlation holds in time in a rather stable way. The slopes we found for our three intervals are steeper than the ones reported in \cite{Li11}. A plausible explanation for such distinction, besides the differences in the dataset, resides in the different spectral models  that we used in this paper; we fitted the energy spectra between 4 and 30 keV with a combination of two power laws, while \cite{Li11} only consider the 3-10 keV energy range.

Following the approach we described in Sec. \ref{timing}, we performed the timing analysis of the entire \lsi{} RXTE/PCA data in a phase-resolved manner, producing average PDS for a set of ten bins, in order to investigate the presence of timing features in the PDS and, particularly, its possible connection with the source periodicity. We did not split the dataset in the three intervals described above in an attempt to improve the (limited) S/N in the PDS. 
The recent discovery of transient radio pulsations by \cite{weng22} implied the possibility of the detection of pulsations or quasi-periodic modulations in the X-ray lightcurves. Nevertheless, despite the high-resolution provided by the RXTE/PCA Good Xenon data, the PDS shown in Fig. \ref{POW} do not exhibit any statistically significant coherent pulsation, QPO nor other type of broad-band component, aside from a very weak red noise component at low frequencies. 
The absence of timing features in the PDS is consistent with previous studies \citep[see e.g.,][a 95 ks Chandra observation, using the ACIS-S camera in continuous clocking mode]{rea10}.

According to the literature, there are three potentially interesting characteristic frequencies to search for timing signals: (I) \cite{weng22} reported the existence of transient radio pulsation from the direction of \lsi{} with a frequency $f=3.715$~Hz (i.e., P$=269.15508 \pm 0.00016$~ms), with a significance $>20\sigma$. This is the first statistically significant evidence of pulsations from the source at any frequency. (II) During a RXTE monitoring of \lsi{} and by a spectral analysis of a period of strong variability, \cite{tel} suggested the existence of a strong red-noise component and an apparent QPO at $f=2$~Hz \citep[see also,][]{massizim10}. Indeed, \cite{smith09} and \cite{Li11} presented a detailed timing analysis of the \lsi{} main flares, showing that one of them seemed to reveal a tentative QPO at $f\sim2$~Hz. Nevertheless, the authors concluded that the feature was not statistically significant. (III) $\mathrm{P}_{\mathrm{spin}}\sim 11$~s ($f\sim0.09$~Hz), which is the frequency of the neutron star predicted by the Corbet diagram \citep{zamanov13} and that seems to be valid for wind-fed sources. 

Given the detection of a pulsation reported in \cite{weng22},  we investigated the PDS feature which is visible in phase $\phi=0.7$, at frequency $~0.08-0.09$~Hz (Fig.\ref{POW}).  The fitting of the PDS does not return any significant narrow component, and therefore we conclude that this feature is possibly the result of statistical fluctuations. 

The rms shows a moderate modulation with the phase, especially for Interval I and Interval II. In Interval III, the rms also shows a tentative phase modulation, but the profile is severely flattened with respect to the first two intervals. As we show in Fig. \ref{results}, the rms maxima coincides with the dips of X-ray flux (and count rate), which is maximum in the first phase bin for all intervals, and interestingly, in those bins that gather the largest flares of the source (see Fig. \ref{flares}). However, the possibility that the rms is modulated by the presence of flares seems unlikely, \JL{since as described in Sec.\ref{timing} outliers are removed by a specific GTI selection}. Our results show that the rms tend to decrease when the flux increases and \textit{vice-versa}, which is typical of accreting systems such as low-mass X-ray binaries, where higher fluxes corresponds to lower variability levels due to the increased contribution of non-variable photons from the optically thick, geometrically thin accretion disc. However, increased fluxes also correspond to softer spectra in such systems. But in the case of \lsi{}, save perhaps for Interval I, higher fluxes correspond to harder spectra (i.e., lower photon index). This behaviour does not necessarily exclude the possibility of an accretion disk around \lsi{}, but suggests that - if present - the disc around the compact object in \lsi{} must have peculiar properties that make it different from the discs usually observed in low-mass X-ray binaries (e.g., it is particularly small compared to the size of the system, and possibly warped).

\section{Summary and Conclusions}
\label{conc}


\JLM{We have analysed all the archival RXTE/PCA X-ray data of \lsi{}, taken between 1996 and 2011, in order to investigate the rapid X-ray time variability of the source. Using the intrinsic period found by timing techniques, we performed the phase-resolved analysis of the data, obtaining a set of phase-bin-averaged energy spectra and power density spectra. In the following, we summarize our main conclusions:
\begin{itemize}
    \item We have searched for the RXTE/PCA intrinsic period using three independent timing techniques: (1) the LS periodogram, (2) the PDM method and (3) amplitude maximization with sinusoid fitting in the time domain. All three methods yielded a period P$\approx 26.6\pm 0.3$~days, compatible within $\sim1\sigma$ with either the orbital period (P$_1$) and the precession period (P$_2$). Thus, this dataset does not allow us to distinguish between them. This result is consistent with previous estimations using RXTE/PCA data.   
    \item The phase-averaged PDS do not show any statistically significant periodic or aperiodic signal, aside from a weak red noise component at low frequencies (i.e., $<0.1$~Hz). The amplitude of such noise component  shows a moderate phase dependence, although no strong correlation along all phases could be found. Our results also show that the rms variability tends to decrease when the flux increases, something which is typical of accreting binary systems. Moreover, in agreement with previous studies, the flux is anti-correlated with the photon index, meaning that higher values of rms are also related with softer spectra. This may indicate the presence of a small accretion disc in the system.
    \item The data show a clear phase modulation when folded on the resolved period. A significant phase shift between the X-ray flux peak of Interval I and Intervals II/III is evident. Such a shift could be possibly explained as an effect of a super-orbital periodicity, which might also underlay the data, albeit not showing clearly enough to be accounted for in the periodograms. 
    \item In the second interval that we considered for the analysis, we found a well-resolved two-peak phase-folded light-curve, which is also hinted on the first one. The absence of a second peak in the last interval suggests that this feature is either transient or variable in amplitude, which at times might causes smearing in the light-curve profile.

    \item Our timing analysis does not shed light on the nature of the system, nor on the type of compact object that powers the binary. Therefore, further observations of \lsi{} in the X-ray energy range are required to investigate the source behavior in deeper detail. 
\end{itemize}}

\section*{Acknowledgements}

The project that gave rise to these results received the support of a fellowship from ”la Caixa” Foundation (ID 100010434). The fellowship code is LCF/BQ/DR19/11740030. J.L.M acknowledges additional support from the Spanish Ministerio de Ciencia through grant PID2019-105510GB-C31/AEI/10.13039/501100011033. We thank the anonymous referee for all the constructive comments and suggestions that helped to improve the quality of the manuscript.

\section*{Data Availability}

The data used in this paper are publicly available in the HEASARC RXTE archive (\url{https://heasarc.gsfc.nasa.gov/docs/xte/archive.html}).The tools used for the data reduction are publicly available as part of the FTOOLS package (\url{https://heasarc.gsfc.nasa.gov/ftools/}). The codes employed for the data analysis presented here will be available upon reasonable request to the corresponding author.



\bibliographystyle{mnras}
\bibliography{example} 

\bsp	
\label{lastpage}
\end{document}